# Tirocinio Formativo Attivo nell'Ateneo di Palermo: Classe di Abilitazione A049 – Matematica e Fisica

**Aurelio Agliolo Gallitto[a] e Lucia Lupo[b]**

[a] Dipartimento di Fisica e Chimica, Università di Palermo
[b] Liceo Scientifico "Galileo Galilei", Palermo

E-mail: aurelio.agliologallitto@unipa.it, lucia.lupo@istruzione.it

**Riassunto.** Sarà descritta l'organizzazione del primo ciclo del Corso di *Tirocinio Formativo Attivo*, Classe A049 – Matematica e Fisica, nell'Ateneo di Palermo, sia per quanto riguarda le attività psico-pedagogiche, sia per le attività disciplinari. In particolare, sarà descritta l'organizzazione del tirocinio svolto a scuola sotto la guida dei tutor accoglienti e sotto la guida del tutor coordinatore. Saranno evidenziati punti di forza, punti di debolezza e aspetti di criticità delle varie attività svolte e verranno proposti miglioramenti per i cicli successivi.

**Abstract.** We will describe the organization of the first cycle of the Course of *Tirocinio Formativo Attivo*, Class A049 – Mathematics and Physics, at the university of Palermo, both in terms of psycho-pedagogical activities and in terms of subject activities. In particular, we will describe the organization of the training done at school under the guidance of welcoming tutors and under the guidance of the coordinating tutor. The strengths, weaknesses and critical aspects of various activities will be highlighted and improvements will be proposed for the subsequent cycles.

## 1. Introduzione

Il reclutamento degli insegnanti nella scuola secondaria, in Italia, è stato attuato fino al 1999 con concorsi per titoli ed esami, nei quali era possibile ottenere il ruolo sul contingente di posti messi a concorso o il titolo di abilitazione che dava l'accesso alle graduatorie provinciali. L'individuazione degli aventi diritto all'immissione in ruolo veniva effettuata attingendo per il 50% dei posti disponibili alle graduatorie relative al concorso (valide fino all'entrata in vigore della graduatoria del concorso successivo), per il restante 50% alle graduatorie provinciali.

Nell'anno accademico 1999/2000, facendo seguito alle indicazioni emerse negli accordi di Lisbona del 1989, le Università avviarono le Scuole di Specializzazione all'Insegnamento Secondario (SSIS), che rimasero attive per nove cicli. Le SSIS abilitavano all'insegnamento, permettendo l'inserimento nelle graduatorie provinciali fino al 2007. Nelle more dell'organizzazione delle Lauree Magistrali per l'insegnamento, sono stati istituiti nel 2010, con il D.M. 249, i Tirocini Formativi Attivi [1], dei quali si parlerà ampiamente nel presente lavoro. Contestualmente, nel 2012, con il D.D.G. 82, vengono indetti nuovi concorsi a cattedra, per titoli ed esami, a cui si accede solo se in possesso del titolo di abilitazione [2]. L'individuazione degli aventi diritto al ruolo resta, al momento, regolata per il 50% alla graduatoria del concorso e il 50% alle graduatorie provinciali permanenti.

Il Tirocinio Formativo Attivo (TFA) è un corso di formazione iniziale per il conseguimento del titolo di abilitazione all'insegnamento nella scuola secondaria di primo e secondo grado. Il TFA, come riporta il D.M. 249/2010, è finalizzato a "*qualificare e valorizzare la funzione docente attraverso l'acquisizione di competenze disciplinari, psico-pedagogiche, metodologico-didattiche, organizzative e relazionali, necessarie a far raggiungere ai futuri insegnanti i risultati di apprendimento previsti dall'ordinamento vigente*". Per i docenti in servizio invece sono stati attivati nel 2004 corsi di formazione nell'ambito del Progetto Lauree Scientifiche [3], convertito nel 2009 in Piano Nazionale Lauree Scientifiche [4].

In questo articolo, descriveremo l'organizzazione del primo ciclo del corso di TFA, Classe A049 – Matematica e Fisica, nell'Ateneo di Palermo, evidenziando punti di forza, punti di debolezza e aspetti di criticità emersi nelle varie attività svolte. Il primo ciclo del TFA parte nell'anno accademico 2011/2012, anche se le attività didattiche sono state svolte nel 2013. Al primo ciclo accedono i candidati che hanno superato le prove di ingresso (test di selezione, prova scritta e orale) e i candidati che, avendo superato le selezioni delle Scuole di Specializzazione nei cicli attivati, non hanno potuto conseguire l'abilitazione perché hanno sospeso la frequenza per completare corsi di Dottorato di Ricerca, i cosiddetti "congelati SSIS". Questi ultimi, al conseguimento dell'abilitazione con il TFA, possono sciogliere la riserva con la quale si sono precedentemente iscritti nelle graduatorie permanenti. A tali graduatorie, definite "a esaurimento" nella legge finanziaria 2007 a firma del ministro Fioroni (D.L. 296/2006), a oggi non hanno accesso coloro che hanno conseguito l'abilitazione iscrivendosi al TFA dopo aver superato l'esame di ammissione [5].

**2. Organizzazione didattica del TFA**

Secondo il D.M. 249, le attività in cui si articola il corso di TFA prevedono l'acquisizione di 60 crediti formativi universitari (CFU), ripartiti secondo le seguenti tipologie.

   a) Insegnamenti di scienze dell'educazione per un totale di 18 CFU, di cui 6 CFU riservati alla didattica e pedagogia speciale rivolte ai bisogni speciali.
   b) Insegnamenti di didattiche disciplinari, per un totale di 18 CFU, svolti nell'ottica di una stretta relazione tra l'approccio disciplinare e l'approccio didattico, anche in un contesto laboratoriale.
   c) Tirocinio diretto e tirocinio indiretto per complessive 475 ore, pari a 19 CFU, da svolgersi presso le istituzioni scolastiche, sotto la guida di un tutor e in collaborazione con il docente universitario relatore della relazione finale di tirocinio, con la quale si conclude l'attività di tirocinio nella scuola; 75 ore del predetto tirocinio sono dedicate alla maturazione delle necessarie competenze didattiche per l'integrazione degli alunni con disabilità.

*2.1 – Insegnamenti di scienze dell'educazione*

Gli insegnamenti di scienze dell'educazione, attivati nell'Ateneo di Palermo, sono riportati in Tabella 1; essi hanno un carattere trasversale alle discipline di insegnamento e forniscono ai tirocinanti le basi per la costruzione delle competenze psico-pedagogiche e metodologico-didattiche.

**Tabella 1.** Insegnamenti di scienze dell'educazione.

| Insegnamento | SSD | CFU |
|---|---|---|
| Metodologia didattica | M-PED/03 | 3 |
| Tecnologie per l'istruzione | M-PED/03 | 3 |
| Pedagogia della scuola | M-PED/01-02 | 3 |
| Valutazione di apprendimenti e competenze | M-PED/04 | 3 |
| Didattica speciale | M-PED/03 | 3 |
| Pedagogia speciale | M-PED/03 | 3 |

I tirocinanti hanno svolto le attività didattiche relative a questa area dalla seconda decade di febbraio 2013 alla seconda decade di aprile 2013, suddivisi in quattro grandi gruppi formati da un centinaio di studenti di differenti classi di abilitazione. Per molti tirocinanti è stato un primo approccio alle tematiche generali relative alla professione docente; il confronto fra laureati di aree disciplinari differenti si è rivelato molto utile e costruttivo. L'esperienza di lavoro cooperativo fra pari, provenienti da percorsi formativi universitari e professionali diversificati, ha permesso ai tirocinanti di acquisire quelle competenze relazionali che sono alla base del buon funzionamento degli organi collegiali nella scuola.

*2.2 – Insegnamenti di didattiche disciplinari*

Gli insegnamenti di didattiche disciplinari, svolti anche in ambiente laboratoriale, mirano a stabilire una stretta relazione tra l'approccio disciplinare e l'approccio didattico. Obiettivo dei suddetti insegnamenti è quello di promuovere nei futuri docenti l'acquisizione di metodologie didattiche finalizzate all'applicazione delle competenze disciplinari nell'attuale contesto scolastico, stabilendo una stretta connessione tra i contenuti disciplinari e l'approccio pedagogico-didattico. In particolare, gli abilitati nella classe A049, alla fine del percorso, devono:

1. possedere la visione della matematica e della fisica come scienze correlate e non a sé stanti: le conoscenze di base delle due discipline devono essere acquisite dagli studenti di scuola superiore, al fine di sviluppare la capacità di elaborare modelli di interpretazione e strumenti culturali, tipici del metodo scientifico e logico-deduttivo, utili nella vita quotidiana per affrontare le sfide del mondo moderno [6];
2. conoscere le Indicazioni Nazionali riguardanti gli obiettivi specifici di apprendimento concernenti le attività e gli insegnamenti compresi nei piani di studio previsti per i percorsi liceali (D.P.R. 89/2010, art. 10, comma 3) [7];
3. acquisire solide conoscenze in didattica della fisica e della matematica ed essere capaci di ricostruire i saperi esperti mettendo in relazione l'analisi dei contenuti disciplinari con le caratteristiche cognitive degli studenti, in modo da saper organizzare sequenze di insegnamento/apprendimento adeguate al livello scolastico [8],[9],[10];
4. essere in grado di programmare percorsi didattici curriculari a lungo termine, individuando i nuclei fondanti disciplinari (di contenuto e metodologici), le propedeuticità e i tempi necessari alla costruzione degli apprendimenti, contestualizzati negli specifici indirizzi scolastici, scegliendo di volta in volta le metodologie e gli strumenti più appropriati al percorso previsto [11],[12].

Nel corso di TFA per la Classe A049 nell'Ateneo di Palermo, sono stati attivati gli insegnamenti disciplinari elencati in Tabella 2.

**Tabella 2.** Insegnamenti disciplinari.

| Insegnamento | SSD | CFU | Docente |
|---|---|---|---|
| Didattica della fisica e innovazioni (***) | FIS/08 | 3 | Claudio Fazio |
| Didattica della matematica e innovazioni (*)(**) | MAT/04 | 3 | Teresa Marino |
| Storia della matematica (*)(**) | MAT/04 | 3 | Cinzia Cerroni |
| Storia della fisica (***) | FIS/08 | 3 | Rosamaria Sperandeo Mineo |
| Laboratorio di didattica della fisica | FIS/01-08 | 3 | Aurelio Agliolo Gallitto |
| Laboratorio di didattica della matematica | MAT/01-08 | 3 | Aldo Brigaglia |

(*) Corso integrato
(**) Corso mutuato con la Classe A047 – Matematica
(***) Corso mutuato con la Classe A038 – Fisica

La struttura dell'offerta formativa è indicativa della scelta operata dal gruppo di progetto, che ha individuato una sezione dedicata alla didattica in una prospettiva innovativa, una sezione dedicata alla storia e un'area dedicata alla riflessione sulla didattica. La matematica e la fisica hanno uno sviluppo storico ed epistemologico connesso, ma allo stesso tempo autonomo. È stato possibile far emergere le peculiarità metodologiche e didattiche delle due discipline attraverso una riflessione attenta condotta in parallelo dai tirocinanti sotto la guida di docenti afferenti ai SSD specifici, nei corsi riguardanti le prime due sezioni.

Il punto di forza di questa sezione è rappresentato dall'opportunità data ai tirocinanti, e da loro pienamente colta, di effettuare un'analisi storico-epistemologica su temi fondanti. Molti tirocinanti, infatti, non avevano avuto modo di approfondire questi aspetti nel loro percorso accademico e hanno mostrato un interesse che avrà modo di svilupparsi nella formazione permanente dei futuri docenti. In una fase successiva (maggio - giugno 2013), i tirocinanti hanno avuto modo di riflettere su quanto acquisito in termini di riflessione storica e metodologie didattiche in un setting di tipo laboratoriale. In entrambi i laboratori di fisica e di matematica, i tirocinanti hanno analizzato alcuni argomenti disciplinari trattati negli insegnamenti proposti, quindi hanno sviluppato dei percorsi innovativi basati sulle attività di laboratorio e sulle problematiche epistemologiche e didattiche precedentemente evidenziate. Il collegamento fra quanto introdotto nelle didattiche e quanto sperimentato nei laboratori è stato molto stretto; per esempio, a partire dal laboratorio di matematica, i tirocinanti hanno avuto modo di sviluppare competenze nell'uso di Geogebra [13], un software *open source* di geometria dinamica, che si è rivelato uno strumento didattico flessibile e utile nella progettazione di percorsi didattici trasversali alle due discipline. I laboratori di didattica hanno rappresentato per tutti i tirocinanti un momento altamente formativo, poiché hanno permesso loro di rielaborare in modo operativo conoscenze teoriche di livello esperto in una prospettiva di insegnamento a studenti di scuola superiore.

**3. Tirocinio e relazione finale**

Le attività di tirocinio prevedono un monte di 475 ore, pari a 19 CFU, svolte in collaborazione con le istituzioni scolastiche del sistema nazionale dell'istruzione sotto la guida di personale docente in servizio nelle istituzioni scolastiche. Le ore sono divise in tirocinio indiretto e tirocinio diretto e gestite dal tutor coordinatore e dai tutor accoglienti. Le 25 ore relative a ogni CFU sono state distribuite, vista l'eccezionalità dei tempi di svolgimento del ciclo, in 8 ore di tirocinio diretto e 17 ore di tirocinio indiretto. I tirocinanti hanno avuto l'opportunità di richiedere la riduzione del carico relativo nel caso avessero prestato servizio presso le istituzioni scolastiche, conseguito il titolo di dottore di ricerca o di master.

Il tirocinio indiretto è stato svolto dai tirocinanti sotto la guida del tutor coordinatore, un docente in servizio selezionato tramite concorso (Decreto Rettorale n. 4729 del 10/12/2012) e utilizzato in regime di tempo parziale. I compiti del tutor coordinatore sono:

- orientare e gestire i rapporti con i tutor accoglienti assegnando gli studenti alle diverse classi e scuole e formalizzando il progetto di tirocinio dei singoli studenti;
- provvedere alla formazione del gruppo di studenti attraverso le attività di tirocinio indiretto e l'esame dei materiali prodotti dagli studenti nelle attività di tirocinio;
- supervisionare e valutare le attività del tirocinio diretto e indiretto;
- seguire le relazioni finali per quanto riguarda le attività in classe.

Il tirocinio diretto è stato svolto presso le istituzioni scolastiche sotto la guida dei tutor accoglienti, docenti in servizio a tempo indeterminato da almeno 5 anni, che sono stati selezionati tramite bando interno dai Dirigenti scolastici delle scuole iscritte nell'elenco di cui all'art. 12 del D.M. 249/2010 e che hanno stipulato una convenzione con l'Università. Ai tutor accoglienti è affidato il compito di:

- orientare gli studenti rispetto agli assetti organizzativi e didattici della scuola e alle diverse attività e pratiche in classe;
- accompagnare e monitorare l'inserimento in classe e la gestione diretta dei processi di insegnamento degli studenti tirocinanti.

*3.1 – Tirocinio diretto a scuola sotto la guida del tutor accogliente*

Per la Classe A049, nell'Ateneo di Palermo, viste le richieste dei tirocinanti e le disponibilità delle istituzioni scolastiche, sono stati incaricati i docenti elencati nella Tabella 3. In alcuni casi, per ragioni organizzative, si è ritenuto necessario affidare il tirocinante a due tutor, nella maggior parte dei casi è stato privilegiato un rapporto uno a uno per facilitare la conoscenza reciproca fra tutor e tirocinante.

I percorsi di tirocinio diretto sono stati sviluppati dai tutor accoglienti tenendo conto delle esigenze dei tirocinanti e concordando con questi sia le attività che i tempi. A tutti i tutor accoglienti è stato richiesto di programmare le attività considerando l'esigenza dei tirocinanti di familiarizzare con la struttura organizzativa della scuola, dal punto di vista logistico e didattico. È stato richiesto di far partecipare i tirocinanti alle attività programmate nella scuola nel periodo di svolgimento del tirocinio, favorendone la presenza alle riunioni degli organi collegiali, di permettere la comprensione dell'uso e della funzione degli strumenti della docenza (formali e valutativi), di comprendere le esigenze di progettazione e documentazione delle attività didattiche. Il percorso doveva inoltre prevedere una fase osservativa del lavoro del docente esperto e una parte attiva nella quale il tirocinante potesse mettere in pratica quanto studiato e osservato, sotto l'attento monitoraggio del docente tutor. Queste linee guida sono state interpretate e messe in pratica dai tutor accoglienti in piena autonomia come previsto dalla funzione docente [14].

**Tabella 3.** Istituzioni Scolastiche e tutor accoglienti.

| Istituzione Scolastica | Tutor |
| --- | --- |
| Liceo Scientifico "Galileo Galilei", Palermo | Cambiaso Nicolò |
| | Falsone Angela |
| | Principato Giacomo |
| | Raimondi Maria Assunta |
| | Rizzo Daniela Carla |
| Liceo Classico "G. Meli", Palermo | Bosco Luigi |
| | Di Lorenzo Concetta |
| | Oddo Lucia |
| | Sutera Diego |
| Liceo Scientifico "S. Cannizzaro", Palermo | Marchisotta Giuseppa |
| | Zanca Antonio |
| IM "Regina Margherita", Palermo | Cordone Giulia |
| | Neri Antonia |
| Liceo Scientifico "Benedetto Croce", Palermo | Ignaccolo Paolo |
| IISF "Ferrara", Palermo | Fascetta Antonino |
| IIS "Picone", Lercara Friddi (Pa) | Di Palermo Cosimo |
| Liceo Scientifico "Don Colletto", Corleone (Pa) | Scalisi Rosa |
| Liceo Scientifico "M. Cipolla", Castelvetrano (Tp) | Spinelli Francesco |
| Istituto "Don Bosco", Palermo | Biondi Alessandro |
| IP "S.M. Mazzarello", Palermo | Masi Rosanna |
| IP Centro Lingue Misilmeri (Pa) | Oliveri Laura |
| Liceo Scientifico "E. Fermi", Agrigento | Mingoia Giovanni |

*3.2 – Tirocinio indiretto sotto la guida del tutor coordinatore*

Durante gli incontri di tirocinio indiretto, sono state proposte le tematiche relative alla funzione docente (il contratto di lavoro, gli strumenti della docenza), alla normativa scolastica alla luce dello sviluppo delle riforme, alla didattica disciplinare e per competenze, alla programmazione e gestione delle attività extrascolastiche (progetti europei, POF, PON e POR) alla valutazione interna ed esterna nella scuola (Sistema Nazionale di Valutazione, valutazione nazionale e internazionale, prove INVALSI, PISA e problematiche connesse) [15]. I tirocinanti hanno contestualmente condotto lo studio individuale, elaborando dei lavori monografici sui temi:

- autonomia scolastica e Piano dell'Offerta Formativa (POF);
- riforma dei cicli/competenze chiave;
- didattica per competenze;
- valutazione di sistema e di istituto;
- percorsi didattici disciplinari.

Durante tutti gli incontri è stato favorito il confronto fra pari e sono state proposte attività che potessero sviluppare atteggiamenti atti al lavoro di gruppo e alla cooperazione; nella fase conclusiva è stata incoraggiata la riflessione critica sull'intero percorso di tirocinio, in vista della predisposizione della relazione finale.

*3.3 – Relazione di tirocinio ed esame finale di abilitazione*

L'attività di tirocinio nella scuola, ̶ come disposto nell'art. 10 comma 6 del D.M. 249 "*si conclude con la stesura da parte del tirocinante di una relazione del lavoro svolto in collaborazione con l'insegnante tutor che ne ha seguito l'attività.[…] La relazione consiste in un elaborato originale che, oltre all'esposizione delle attività svolte dal tirocinante, deve evidenziare la capacità del medesimo di integrare ad un elevato livello culturale e scientifico con le competenze acquisite nell'attività svolta in classe e le conoscenze in materia psico-pedagogica con le competenze acquisite nell'ambito della didattica disciplinare e, in particolar modo, nelle attività di laboratorio*". I tirocinanti hanno redatto l'elaborato, su indicazione del gruppo di lavoro, in tre parti. Nella prima parte è stata descritta l'esperienza formativa all'interno della scuola, in collaborazione con il tutor accogliente; nella seconda parte è stato rielaborato quanto approfondito durante il tirocinio indiretto; nella terza parte il tirocinante ha rielaborato un'esperienza significativa legata all'area della didattica e dei laboratori, applicando le conoscenze psico-pedagogiche acquisite e riflettendo, alla luce dell'esperienza diretta nella scuola, sulla spendibilità dell'esperienza in ambito didattico. Il relatore è stato il docente universitario che ha proposto durante il suo corso l'esperienza significativa scelta dal tirocinante, correlatore l'insegnante tutor coordinatore. Gli argomenti trattati dai tirocinanti nella terza parte della relazione sono stati rielaborati dai tirocinanti stessi in forma di articolo e sono stati raccolti nel presente numero della rivista.

Al termine del percorso di tirocinio si è svolto l'esame finale di abilitazione, che è consistito nella valutazione da parte della commissione (nominata con Decreto Rettorale n. 1853/2013 del 20/06/2013) dell'attività svolta durante il tirocinio, sulla base degli elementi forniti dalle schede valutative redatte dai tutor scolastici, dall'esposizione orale di un percorso didattico su un tema scelto dalla commissione, dalla discussione della relazione finale di tirocinio. I temi assegnati dalla commissione sono stati sviluppati e quindi discussi dai tirocinanti, seguendo le indicazioni fornite dalla commissione stessa: "*Il candidato tracci un percorso didattico, che coinvolga matematica e fisica, specificando la classe a cui è rivolto. Il candidato deve specificare prerequisiti, obiettivi, metodologie e strumenti di verifica e presentare i contenuti con un learning object*". L'elenco dei temi assegnati è riportato qui di seguito.

- Funzioni periodiche in matematica e fisica
- Vettori e loro applicazioni fisiche
- I logaritmi, la funzione esponenziale: scarica del condensatore
- Il problema della tangente, la derivata e applicazioni cinematiche
- La trigonometria nel piano inclinato
- Il problema dell'area, l'integrale definito e il lavoro compiuto da una forza
- La funzione integrale e il lavoro compiuto da una forza dipendente dalla posizione: le forze elastiche
- La cicloide e il problema dell'isocronismo del pendolo
- Le funzioni goniometriche e l'oscillatore armonico semplice: il sistema massa molla
- Linearità, proporzionalità diretta e applicazioni fisiche
- Gli errori casuali nelle misure fisiche e la funzione di Gauss
- Le orbite dei pianeti
- Prodotto vettoriale e momento angolare
- Prodotto scalare e lavoro compiuto da una forza
- Moto del proiettile e traiettorie paraboliche
- Proporzionalità inversa, equazione dell'iperbole e legge dei gas perfetti
- Sistemi di riferimento polari e moto circolare uniforme
- Similitudini e leggi di scala
- La similitudine dei triangoli nella scomposizione delle forze
- La misura di pi-greco: metodi matematici e metodi fisici
- Le simmetrie in natura

## 4. Considerazioni conclusive

Le attività didattiche del primo ciclo di TFA sono iniziate con un notevole ritardo rispetto ai tempi previsti dalla legge, determinando una situazione di emergenza nella quale tutti gli attori hanno dovuto dare il massimo per l'ottimizzazione dei tempi con l'obiettivo, pienamente raggiunto, di completare entro pochi mesi il percorso formativo nella sua interezza e complessità.

Sempre a causa dei tempi ristretti, si è determinato altresì uno sfalsamento delle attività relative al tirocinio e ai laboratori pedagogico-didattici; questi, infatti, sono indirizzati alla rielaborazione e al confronto delle esperienze di tirocinio. Praticamente, è stato impossibile coordinare in poche settimane, alla fine dell'anno scolastico, una progettazione condivisa fra tutor e docenti universitari. Per i prossimi cicli si auspica l'attivazione dei Consigli di Tirocinio, nell'ambito dei quali si possa discutere della progettazione delle suddette attività.

Infine, va segnalata, come elemento di criticità, la mancanza di un adeguato approfondimento storico-didattico di alcune tematiche relative alle due discipline; ad esempio, per la matematica, probabilità e statistica che sono presenti in maniera consistente nelle indicazioni nazionali e sono considerati dall'Unione Matematica Italiana (UMI) nucleo fondante; per la fisica, ottica e fisica moderna. Per questi temi, visti i tempi ristretti, non si è avuto modo di progettare un momento di riflessione ma nella logica di una selezione di fondamenti si è operata questa scelta considerando prioritaria la costruzione di competenze su contenuti ridotti piuttosto che la trattazione di molti contenuti a scapito della costruzione di competenze.